# A facility for mass production of ultra-pure NaI powder for the COSINE-200 experiment

**Keonah Shin,**[a] **Junseok Choe,**[a] **Olga Gileva,**[a] **Alain Iltis,**[b] **Yeongduk Kim,**[a,c]
**Cheolho Lee,**[a] **Hyun Su Lee,**[a,c*] **Moo Hyun Lee,**[a,c] **and Hyang Kyu Park**[d]

[a] *Center for Underground Physics, Institute for Basic Science (IBS),*
   *Daejeon, 34126, Korea*

[b] *Damavan Imaging,*
   *Troyes, 10430, France*

[c] *IBS School, University of Science and Technology (UST),*
   *Daejeon, 34113, Korea*

[d] *Department of Accelerator Science, Korea University,*
   *Sejong, 30019, Korea*

   E-mail: hyunsulee@ibs.re.kr

ABSTRACT: COSINE-200 is the next phase of the ongoing COSINE-100 experiment. The main purpose of the experiment is the performance of an unambiguous verification of the annual modulation signals observed by the DAMA experiment. The success of the experiment critically depends on the production of a 200 kg array of ultra-pure NaI(Tl) crystal detectors that have lower backgrounds than the DAMA crystals. The purification of raw powder is the initial but important step toward the production of ultra-pure NaI(Tl) detectors. We have already demonstrated that fractional recrystallization from water solutions is an effective method for the removal of the problematic K and Pb elements. For the mass production of purified powder, a clean facility for the fractional recrystallization had been constructed at the Institute for Basic Science (IBS), Korea. Here, we report the design of the purification process, material recovery, and performance of the NaI powder purification facility.

---

[*] Corresponding author.

# Contents

## 1. Introduction

Various astronomical observations conclude that most of the matter in the Universe is comprised of invisible dark matter, but its nature has remained elusive [1,2]. A previously unseen weakly interacting massive particle (WIMP) [3,4] has been proposed as a possible explanation for the dark matter phenomenon. Even though numerous experiments have searched for particle dark matter candidates, with the notable exception of the DAMA experiment, none have observed a definitive signal [5,6]. The DAMA experiment observes an annual modulating signal event rate in an array of NaI(Tl) detectors [7,8] that can be interpreted as a result of WIMP-nucleon interactions [9,10]. However, there have been long-standing questions about this interpretation because no other experimental searches have observed similar signals during the past 20 years [11].

    The COSINE experiment aims to reproduce the DAMA's annual modulation signal using the same NaI(Tl) target materials [12,13,14]. Since September 2016, COSINE-100 has operated a 106 kg array of low-background NaI(Tl) crystals at the Yangyang underground laboratory in Korea. Recent physics results from the COSINE-100 experiment [15,16,17,18] could not reach an unambiguous conclusion because of higher crystal background rates than those of DAMA. Consequently, COSINE-200, which will use a 200 kg NaI(Tl) crystal array with a similar or lower background level than those in the DAMA experiment, has been proposed. For the development of ultra-pure NaI(Tl) detectors, all steps in the production of low-background detectors, which include raw powder purification, ultra-pure NaI(Tl) crystal growth, and the detector assembly, have been carefully re-examined and improved.

    The COSINE-200 detector requires extremely low levels of radioactive contamination in the materials used in the detector production. The major contributors to the background are the decays of $^{40}$K and $^{210}$Pb in the NaI(Tl) crystal bulk [14,19]. Because of the similarity of their chemical properties with those of Na, which is in the same group of the periodic table, K is the main impurity contaminant and its selective extraction from the NaI powder is challenging. A few

methods have been applied to purify the NaI powder, including ion-exchange column chromatography with Pb and Ra resins [20] and a fractional recrystallization method [21]. In a laboratory-scale experiment, fractional recrystallization was found to be effective for the reduction of both K and Pb impurities. In addition, the Ba concentration was significantly reduced, which indicates a reduction of Ra impurities.

Based on the success of the laboratory-scale experiments, we assembled a mass production facility for the ultra-pure NaI powder with the fractional recrystallization method on-site at the Institute for Basic Science (IBS), Daejeon, Korea. The facility is designed to produce approximately 35 kg of ultra-pure powder in a single processing cycle. In this report, we describe the mass production facility for purification of the NaI powder that will be used for ultra-pure NaI(Tl) crystal growth [22] for the COSINE-200 experiment, including a description of the apparatus and its initial purification performance.

## 2. Facility and experimental conditions

### 2.1 Mass production facility

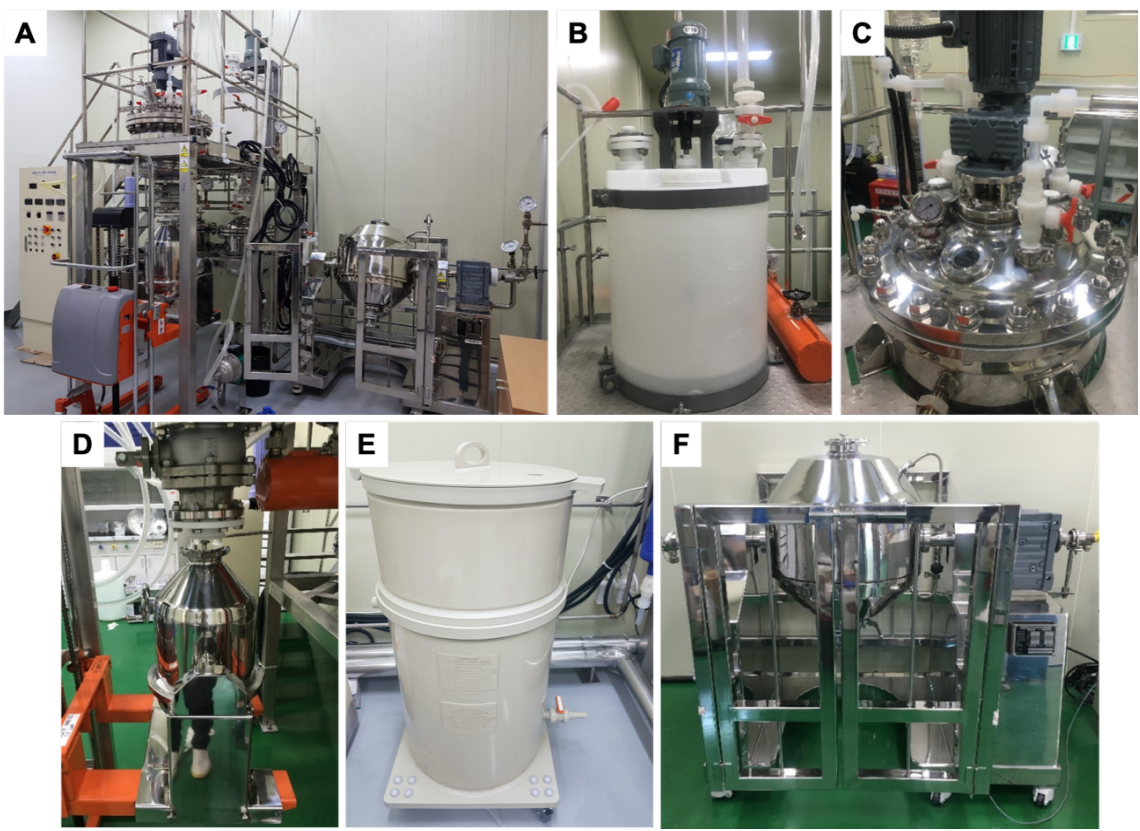

**Figure 1**. The custom-made recrystallization facility for mass production of the ultra-pure NaI powder. A: Overall picture of the facility, B: Dissolving tank, C: Recrystallization tank, D: Initial filter housing, E: Nutsche filter housing, F: Biconical vacuum rotary dryer.

Pictures of the custom-made NaI powder purification facility at IBS are shown in Fig. 1. The facility consists of a dissolving tank, a recrystallization tank, a filter housing, and a biconical vacuum rotary dryer. The design value of the maximum charge of initial NaI powder is 70 kg, and the target recovery rate is approximately 50%. Therefore, one expects to produce approximately 35 kg of the purified NaI powder in a single batch cycle. The entire process for a single batch cycle takes about five working days.

The dissolving tank (Fig. 1-B) is comprised of polyethylene (PE) material in a cylindrical shape with a total volume of 120 L. It has a large inlet on the top (160 mm diameter) that allows for easy charging of the initial powder and de-ionized (DI) water. Because of the highly corrosive and abrasive properties of the NaI sludge, a titanium agitator is used to mix and dissolve the powder in the DI-water. The agitator is controlled by a programmable motor system. A Teflon tube with a ball valve connected to the bottom of the tank serves as the outlet for the saturated NaI solution and its connection to the recrystallization tank (Fig. 1-C). To remove insoluble impurities, a polypropylene (PP) cartridge filter is installed in the middle of the outlet tube.

The recrystallization tank is a stainless-steel tank equipped with an oil jacket that controls its temperature. A perfluoroalkoxy alkane (PFA) lining covers all of the stainless-steel tank's internal surfaces for protection against corrosion caused by the NaI sludge. A temperature control unit (TCU) heats up and circulates the silicon oil inside the jacket. The maximum capacity of the recrystallization tank is 50 L solution. A viewport on the top provides the ability to observe the amount of solution, any colour changes, and recrystallization status. It is connected to an acid-resistant air diaphragm pump that produces the vacuum. The vacuum pump transfers the NaI solution from the dissolving tank into the recrystallization tank. It also removes water vapor produced during the evaporation process. A larger ball valve (150 mm diameter) is located at the bottom of the tank for the extraction of the produced sludge of the recrystallized NaI along with the mother solution.

To separate the recrystallized material, the NaI sludge from the recrystallization tank is dropped into the filter housing through the bottom ball valve. Initially we used a special filter housing that was designed to make a direct connection with the recrystallization tank without air contact (See Fig. 1-D). This had the advantage of minimizing the contamination from the outside environment. However, the relatively small inner diameter (300 mm) made a long processing time (a few hours) for the filtration and washing of a ~ 20 kg charge of wet NaI. This was mainly because the collected crystals caked and clogged the filter. We searched for a commercially available alternative option and found a nutsche filter (Fig. 1-E) from the Russian company "Rusredmet" that is made of polypropylene (PP) material.

The nutsche filter housing has two divided sections, one for collecting crystals and another for receiving the mother solution. The volume of the collecting and receiving parts are 40 L and 80 L, respectively. The receiving part has two outlets, one for a vacuum line and the other for the removal of the mother solution. The 500 mm diameter allows for a fast filtration process. After washing with alcohol, the filtered crystals are transferred to the rotary dryer (Fig. 1-F).

The biconical vacuum rotary dryer is used to remove water from the filtered NaI crystals. It is made of stainless-steel with a PFA lining that covers all of the dryer's internal surface. Similar to the recrystallization tank, the dryer is equipped with an oil jacket and TCU for temperature control. Titanium rod tubing is used as a vacuum line during the drying process and for purging with $N_2$ gas to relieve the vacuum after the drying process is complete. The maximum charge of wet NaI crystals in the dryer is about 60 kg (30 L volume). A slow rotation of the dryer mixes the crystals and prevents the formation of large clumps during the drying process.

## 2.2 Reagents and equipment

Crystal-grade NaI powder from the Merck company (99.99% purity grade) was used as the initial material for the purification process. High purity DI-water with 18.2 MΩ·cm resistivity was the main solvent. A hydroxylamine solution NH$_2$OH (50 weight% in water; 99.999% purity grade) and hydriodic acid HI (57weight%; 99.999% purity grade) from the Sigma-Aldrich company were used as pre-treatment reagents. Absolute ethanol (~200 proof, HPLC/spectrophotometric grade) from the Sigma-Aldrich company was used to wash the recrystallized NaI crystals. Hydrophilic PTFE membrane filters from the Advantec company with 1 μm pore size were used to separate the recrystallized NaI from the mother solution. Plastic beakers, scoops, and spoons were used instead of glass in order to avoid K contamination. Prior to experiments, all the facilities (tanks and tubes) were cleaned with 2% nitric acid at 80 °C temperature and followed by washing with DI-water three or four times. The concentrations of impurities in the initial and purified samples were measured with an inductively coupled plasma mass spectrometer (ICP-MS) [23]. The water content in the powder was measured with a Karl-Fisher titrator.

### 2.3 Recrystallization procedure

Initially, the required amount of DI-water was introduced into the dissolving tank at room temperature. Then, NaI powder was charged in the same tank while continuously stirring until a saturated solution was reached. To prevent oxidation of iodide-ions, NH$_2$OH was added to the NaI solution with NH$_2$OH : NaI mole ratio of 0.00024 : 1. The solution was stirred for about an hour. Then HI was added until the pH reached 3.5. This solution was then transferred to the recrystallization tank through the tube while maintaining a vacuum. A PP cartridge filter in the middle of the tube was used to separate insoluble impurities. The solution in the recrystallization tank was heated until about 40% of the initial water was evaporated under vacuum pressure. The solution was then slowly cooled down with stirring under vacuum while the crystallization of NaI occurred. By the time it reached room temperature, approximately 50% of the initial NaI powder formed as NaI crystals while the other 50% remained in the mother solution.

The mixture of crystals and the mother solution (sludge) were separated by means of the PTFE membrane filter in the filter housing. The mother solution was collected and recycled for use in the next purification cycle. The filtered crystals were washed with cold absolute ethanol a few times in order to remove the residual mother solution. The purified NaI crystals were transferred to the rotary dryer to remove the crystalline water. The NaI was first dried at 65 °C under vacuum to produce a monohydrate form, then the temperature was raised to 130 °C for the production of anhydrous NaI powder.

### 3. Results and discussion

To study impurity contaminations in the recrystallized NaI, initial crystal-grade NaI powder and purified NaI powder were measured with ICP-MS as summarized in Table 1. Several radioactive contaminants were removed by the recrystallization. In particular, the levels of the two most dangerous contaminants in the NaI(Tl) crystals, K and Pb, were reduced by approximately an order of magnitude. For Sr and Ba, which were used as tracers for Ra, significant reductions were also observed.

The 1$^{st}$ experiment was carried out with a 20 kg initial powder charge. Even though the initial powder contained 248 ppb of K, the recrystallized wet powder contained less than 16 ppb, which satisfies our requirement for the COSINE-200 experiment. However, during the drying process, the powder was contaminated as can be seen in Table 1. It was found that the rotary dryer had a leak of the silicon oil that was used for heating. This was fixed for the 2$^{nd}$ experiment.

**Table 1** ICP-MS results for the initial powder and the purified powders before and after drying for three different experiments. The uncertainties of the ICP-MS measurements are 10%.

| | Sample | K (ppb) | Sr (ppb) | Ba (ppb) | Pb (ppb) | Th (ppt) | U (ppt) |
|---|---|---|---|---|---|---|---|
| | Initial NaI powder | 248 | 19.0 | 2.9 | 40.0 | <6 | <6 |
| 1st | Recrystallized and washed NaI | <16 | 0.4 | 0.2 | <0.3 | <3.3 | <3.5 |
| | Final product | 36 | 1.6 | 2.4 | 28.6 | 8.3 | <3.5 |
| 2nd | Recrystallized and washed NaI | 47 | 2.9 | 0.8 | 5.9 | <3.0 | <4.0 |
| | Final product | 46 | 3.4 | 1.3 | 4.7 | <3.0 | <4.0 |
| 3rd | Recrystallized and washed NaI | <20 | 0.9 | 0.2 | 3.2 | <3.6 | <3.1 |
| | Final product | <20 | 0.6 | 2.5 | 2.3 | <3.6 | <3.1 |

In the 2nd experiment, the initial powder loading was increased to 40 kg. In this experiment, larger contaminations of K were observed in the wet and final dried powders. This was diagnosed as from insufficient separation of the mother solution in the filter housing. Figure 2 shows the wet powder right after the filtration in the original filter housing during the ethanol washing steps. As shown in Fig. 2 (A), the initial filter cake was yellowish because of residual mother solution. After three ethanol washes, pure white powder was observed on top of the filtered crystals (see Fig. 2 (B)). However, when we removed the upper portion of the white powder, we saw evidence of imperfect washing inside as shown in Fig. 2 (C). Since the shape of original filter housing was narrow and long, the wet crystals compressed there and clogged the filter. Even though two ethanol washes were enough for the 10 kg experiment, five washes in the 20 kg experiment were not sufficient for removing the mother solution, as shown in Fig. 2 (C).

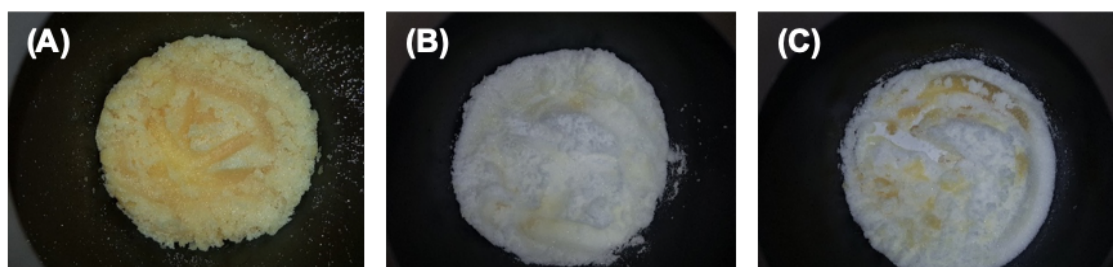

**Figure 2.** Filtration and washing steps with the initial filter housing. (A): Filtered crystals from mother liquor under vacuum. (B): Three times washed crystals with pure ethanol. (C): Upper part of filter cake was removed and the remnants were washed two more times.

The dimensions of the crystal collection part in the original filter housing were 300 mm diameter and 400 mm height. To understand the achievable product purity, a laboratory Büchner funnel was used for the filtration and crystal wash in the 3rd experiment. The initial charge and crystallization conditions were almost the same as those in the 2nd experiment. As shown in Table 1, we achieved a K level that is below 20 ppb. This confirms that the original filter housing was not suitable for more than 10 kg of powder purification.

We have searched for a commercially available alternative option for a large capacity PP filter housing and found a nutsche filter (Fig. 1-E) that has 500 mm diameter from the Russian company "Rusredmet". A large filter surface makes fast filtration and easy ethanol washing.

After three recrystallization experiments, 65 kg of the separated mother solution (containing ~ 41 kg of NaI powder) was collected and used for recovery of the NaI powder. The solution was processed in the same way as the freshly prepared NaI solution. For this experiment, the nutsche filter housing was used. As shown in Fig.3, the recovered NaI powder was washed three times and sufficiently clean NaI powder was obtained.

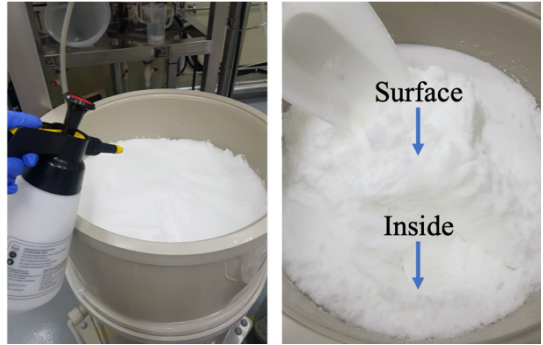

**Figure 3.** Two photos show the filtration and washing step by the nutsche filter housing.

The powder purity is summarized in Table 2. After recrystallization impurities were concentrated in the mother solution; the initial contamination of various radioisotopes was increased by about a factor of two. However, a single recrystallization process followed by filtration with a nutsche filter reduced these impurities by an order of magnitude, a level that satisfies the COSINE-200 requirements.

**Table 2.** Impurities concentration of the initial mother solution and the recovered NaI products. The uncertainties are the same as those in Table 1.

| Samples | K (ppb) | Sr (ppb) | Ba (ppb) | Pb (ppb) | Th (ppt) | U (ppt) |
|---|---|---|---|---|---|---|
| Initial - mother solution (normalized to NaI content) | 565 | 29.3 | 5.1 | 53.9 | <4.3 | <2.6 |
| Final product | <51 | 0.5 | 0.4 | 7.3 | <6.4 | <4.0 |

The water content of the dried powder was measured with a Karl-Fisher titrator. All measurements showed sufficiently low humidity levels of 240 ~ 1000 ppm, which satisfy our goal of less than 1000 ppm.

## 4. Conclusion

A mass production facility for ultra-pure NaI(Tl) powder was constructed at IBS, Korea for the COSINE-200 experiment. A few issues were resolved during the engineering operation. The initial operation demonstrated a relatively fast process for mass purification of NaI powder that successfully reached our goal of a K level below 20 ppb. Based on these results, we will proceed with the mass production of ultra-pure NaI powder for the COSINE-200 experiment.

## Acknowledgments

This work is supported by the Institute for Basic Science (IBS) under project code IBS-R016-A1.

## References

[1] D. Clowe et al., *A direct empirical proof of the existence of dark matter*. *Astrophys. J.* **648** (2006) L109.

[2] N. Aghanim et al., (Planck Collaboration), *Planck 2018 results. VI. Comsmological parameters*. arXiv:1807.06209 [astro-ph.CO].

[3] B. W. Lee and S. Weinberg, *Cosmological lower bound on heavy-neutrino masses*. *Phys. Rev. Lett.* **39** (1977) 165.

[4] M. W. Goodman and E. Witten, *Detectability of Certain Dark Matter Candidates*. *Phys. Rev. D* **31** (1985) 3059.

[5] T. M. Undagoitia and L. Rauch, *Dark matter direct-detection experiments*. *J. Phys. G* **43** (2016) 013001.

[6] M. Schumann, *Direct Detection of WIMP Dark Matter: Concepts and Status*. *J. Phys. G* **46** (2019) 103003.

[7] R. Bernabei et al., (DAMA/LIBRA Collaboration), *Final model independent result of DAMA/LIBRA-phase1*. *Eur. Phys. J. C* **73** (2013) 2648.

[8] R. Bernabei et al., (DAMA/LIBRA Collaboration), *First Model Independent Results from DAMA/LIBRA-Phase2*. *Nucl. Phys. At. Energy* **19** (2018) 307.

[9] C. Savage, G. Gelmini, P. Gondolo, and K. Freese, *Compatibility of DAMA/LIBRA dark matter detection with other searches*. *JCAP* **04** (2009) 010.

[10] Y. J. Ko et al., (COSINE-100 Collaboration), *Comparison between DAMA/LIBRA and COSINE-100 in the light of Quenching Factors*. *JCAP* **11** (2019) 008.

[11] M. Tanabashi et al., (Particle Data Group), *Review of Particle Physics*. *Phys. Rev. D* **98** (2018) 030001.

[12] P. Adhikari et al., (KIMS Collaboration), *Understanding internal backgrounds in NaI(Tl) crystals toward a 200 kg array for the KIMS-NaI experiment*. *Eur. Phys. J. C* **76** (2016) 185.

[13] G. Adhikari et al., (COSINE-100 Collaboration), *Initial Performance of the COSINE-100 Experiment*. *Eur. Phys. J. C* **78** (2018) 107.

[14] P. Adhikari et al., (COSINE-100 Collaboration), *Background model for the NaI(Tl) crystals in COSINE-100*. *Eur. Phys. J. C* **78** (2018) 490.

[15] G. Adhikari et al., (COSINE-100 Collaboration), *An experiment to search for dark-matter interactions using sodium iodide detectors*. *Nature* **564** (2018) 83.

[16] G. Adhikari et al., (COSINE-100 Collaboration), *Search for a dark matter-induced annual modulation signal in NaI(Tl) with the COSINE-100 experiment*. *Phys. Rev. Lett.* **123** (2019) 031302.

[17] C. Ha et al., (COSINE-100 Collaboration), *First Direct Search for Inelastic Boosted Dark Matter with COSINE-100*. *Phys. Rev. Lett.* **122** (2019) 131802.

[18] G. Adhikari et al., (COSINE-100 and Sogang Phenomenology), *COSINE-100 and DAMA/LIBRA-phase2 in WIMP effective models. JCAP* **06** (2019) 048.

[19] G. Adhikari et al., (KIMS Collaboration), Understanding NaI(Tl) crystal background for dark matter searches. *Eur. Phys. J. C* **77** (2017) 437.

[20] K.-I. Fushimi et al., *High purity NaI (Tl) scintillator to search for dark matter. JPS Conf. Proc.* **11**, (2016) 020003.

[21] K. Shin, O. Gileva, Y. Kim, H. S. Lee, and H. Park, *Reduction of the radioactivity in sodium iodide (NaI) powder by recrystallization method. J. Radioanal. Nucl. Chem.* **317** (2018) 1329.

[22] B.J. Park et al., *Development of ultra-pure NaI(Tl) detector for COSINE-200 experiment.* arXiv:2004.06287 [physics.ins-det].

[23] M.H. Lee, *Radioassay and Purification for Experiments at Y2L and Yemilab in Korea*, September, 9-13, 2019 Toyama, Japan. *J. Phys.: Conf. Ser.*, **1468** (2020) 012249.